\documentclass[showpacs,prl,aps]{revtex4}

\usepackage{graphicx}% Include figure files
\usepackage{dcolumn}% Align table columns on decimal point
\usepackage{bm}% bold math
\begin{document}
\title{Entropy scaling laws for diffusion}
\author{Sorin Bastea}
\email{bastea2@llnl.gov}
\affiliation{Lawrence Livermore National Laboratory, P.O. BOX 808, Livermore, CA 94550}
%\date{\today}
\pacs{66.10.Cb}
\maketitle

Recently, Samanta et al. \cite{smag} set out to 
understand the low density failure of the entropy scaling law for the self-diffusion 
coefficient $D$ conjectured by Dzugutov \cite{md96} and to provide a simple 
alternative. After an interesting derivation, that however contains a number of 
uncontrolled approximations, they arrive at Eq. 7 of \cite{smag}, which reduces 
for a hard sphere fluid to: 
\begin{eqnarray}
\frac{D}{D_E}=\frac{A}{1-s_e/k_B}
\end{eqnarray}
with $D_E=D_B/\chi$ and $D_B=3(k_B T/\pi m)^{\frac{1}{2}}/8\rho\sigma^2$ 
the Enskog and Boltzmann diffusion coefficients \cite{cc}, $\chi$ the contact value of the 
pair correlation function, $s_e$ the excess entropy per particle, and $A=2.5$. 
Using the excess entropy $s_e/k_B=-(4\eta-3\eta^2)/(1-\eta)^2$ and contact 
value $\chi=(2-\eta)/2(1-\eta)^3$ given by 
the Carnahan-Starling equation of state \cite{hm} ($\eta=\pi\rho\sigma^3/6$ - 
packing fraction), we test this relation against the molecular dynamics simulation results of 
Erpenbeck and Wood (EW)\cite{ew91} - Fig. 1. 
The disagreement is quite significant and more important the behavior 
of the two curves is very different, as ${(D/D_E)}_{EW}$ is not a monotonically 
decreasing function of $(-s_e/k_B)$. The discrepancy cannot be attributed to the 
authors use of an approximation for the excess entropy.

The idea of entropy scaling for transport coefficients has a fairly 
long history \cite{yr77}. Arguing on the basis of the molecular ``caging'' effect 
Dzugutov \cite{md96} proposed the scaling law:
\begin{eqnarray}
\frac{D}{\sigma^2 \Gamma_E}=B\exp{(s_{e}/k_B)}
\label{eq:dif1}
\end{eqnarray}
which assumes that the natural length and time scales for diffusion are given 
by a suitably defined hard sphere diameter $\sigma$ and the Enskog collision 
frequency $\Gamma_E=4\sigma^2 \chi\rho\sqrt{\pi k_B T/m}$, with $B$ 
an universal constant. Unfortunately, the above relation appears to work only in a 
limited, high density domain for both hard spheres \cite{cr00} and 
realistically modeled fluids \cite{sb03}. The problem that arises at low and moderate 
densities (see Fig. 1) with the scaling introduced in Eq. 2 can be understood if we 
observe that the left-hand-side of that equation can be written up to a multiplicative 
constant as $D/D_B \chi\eta^2$. Therefore, in the limit of a dilute system, $\eta\rightarrow 0$, 
this term diverges as $1/\eta^2$, while the right-hand-side of Eq. 2 remains finite. 
This behavior should be expected for any valid definition of $\sigma$ and $\chi$ and 
$s_{e}$ approximation. 
The noted pathology of Eq. 2 can be avoided by replacing $\sigma$ as the 
relevant length scale with $1/\rho\sigma^2$, the Boltzmann mean-free path, which should 
be a reasonable measure of the degree of molecular confinement, 
$1/\rho\sigma^2\propto \sigma/\eta$. The new relationship is:
\begin{eqnarray}
\frac{D}{D_B \chi}=\exp{(\gamma s_{e}/k_B)}
\end{eqnarray}
where we introduced a different constant $\gamma $. The test of this 
dependence is shown in Fig. 1 for hard spheres, with $\gamma=0.8$. Furthermore, 
Eq. 3 holds for Van der Waals fluids as well \cite{sb03}. 

Samanta et al. also propose a generalized Stokes-Einstein relation. 
However, such a relation is hardly necessary given that the usual Stokes-Einstein 
formula with the 'slip' boundary condition holds well for both hard spheres \cite{agw} 
and Van der Waals fluids \cite{sb03}.

This work was performed under the auspices of the U. S. Department of Energy by 
University of California Lawrence Livermore National Laboratory under Contract 
No. W-7405-Eng-48.

\begin{figure}
\includegraphics{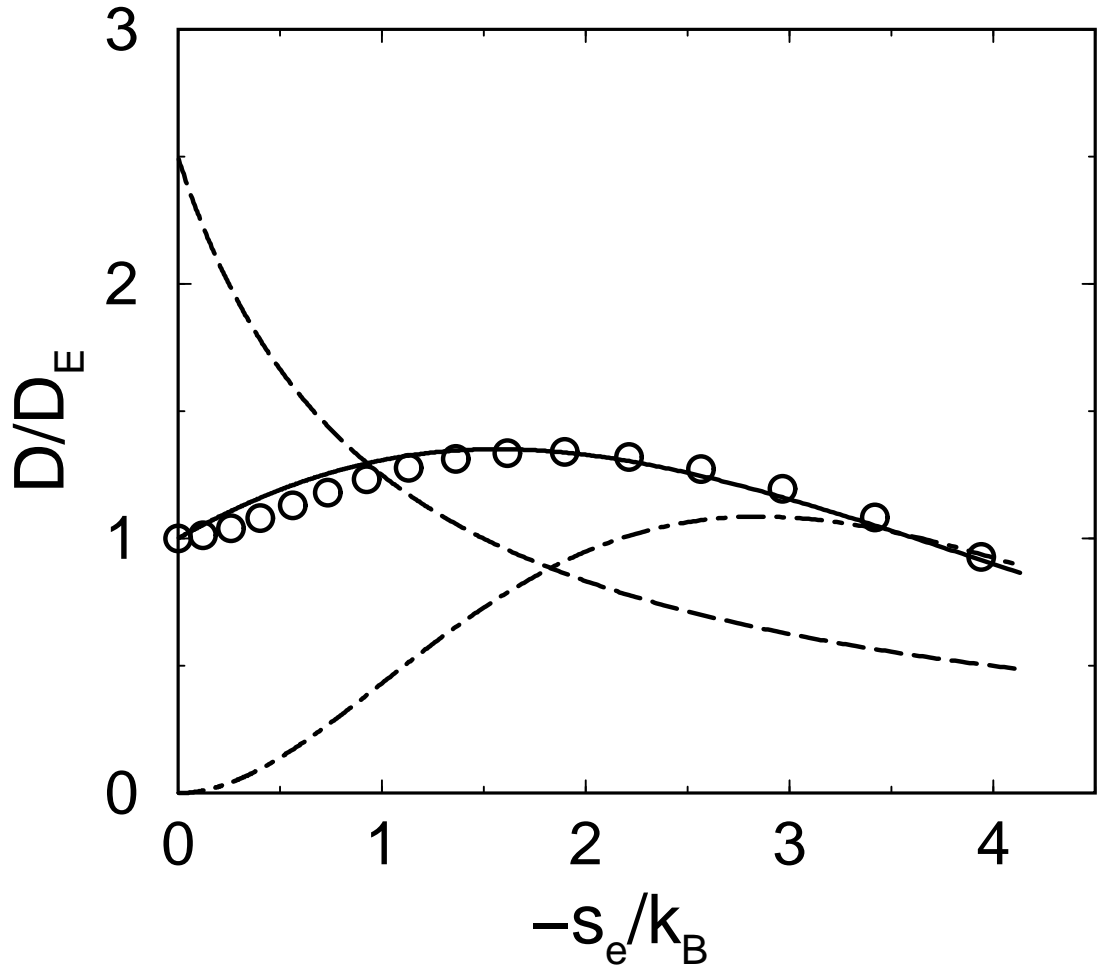}
\caption{Comparison of the hard spheres diffusion coefficient (Ref. \cite{ew91}) 
- circles, with scaling relation of Ref. \cite{smag} (Eq. 1) - dashed line, 
Dzugutov scaling law \cite{md96} (Eq. 2) - dot-dashed line, 
and new entropy scaling (Eq. 3) - solid line.}
\end{figure}
\end{document}